\documentclass[11pt]{article}

\usepackage[margin=1in]{geometry}
\usepackage{amsmath}
\usepackage{amssymb}
\usepackage{authblk}

\usepackage{textcomp}
\usepackage{booktabs}
\usepackage{graphicx}
\usepackage{xcolor}
\usepackage{hyperref}

\title{The Usefulness Gap in Proof-of-Useful-Work: An Empirical Study of Pearl's cuPOW Protocol}

\author{Abhinaba Basu}

\affil{National Institute of Electronics and Information Technology (NIELIT), New Delhi, India}
\affil{Indian Institute of Information Technology Allahabad (IIITA), Prayagraj, India}
\affil{Corresponding author: \texttt{mail@abhinaba.com} \quad ORCID: 0000-0003-1575-4107}

\date{June 2026}

\begin{document}
\sloppy
\maketitle

\begin{center}\small\textit{Preprint. Under review.}\\
Code: \url{https://github.com/abhinaba/pearl-usefulness-gap}\end{center}\vspace{10pt}

\begin{abstract}
Pearl, a Layer-1 blockchain with high-profile AI industry endorsements, markets its Proof-of-Useful-Work (PoUW) protocol as simultaneously securing the network and performing AI inference. We present the first systematic empirical measurement of a deployed PoUW system, finding that Pearl's 24~EH/s network---representing $\sim$320,000 GPU-equivalents consuming an estimated 112~MW---produces zero useful AI computation. Budget GPU rental prices rose 38\% and utilization surged from 57\% to 94\% following the mining software's public release, displacing legitimate research workloads.

Our measurements span five dimensions: (1)~network composition analysis of 8,012 workers shows all have inference-capable hardware, yet the dominant mining software contains no inference code; (2)~the verification protocol accepts random matrices by design, confirmed by 44~pool-accepted shares from our open-source miner across NVIDIA, AMD, CPU, and Apple Silicon hardware; (3)~statistical distribution checks are trivially defeated by adversarial Gaussian sampling; (4)~mining economics are marginal at current PRL prices (\$0.76), with ROI ranging from $-1\%$ to $+67\%$ depending on GPU tier---near breakeven for most hardware; and (5)~the mining computation is commodity integer arithmetic portable to any hardware platform, offering no vendor lock-in. These findings quantify the verifiability-usefulness tension identified theoretically by Hoffmann, providing concrete measurements of its magnitude and economic consequences in a deployed system.
\end{abstract}

\section{Introduction}

Budget GPU rental prices on vast.ai rose 38\% in two weeks following the public release of Pearl's mining software in May~2026, with utilization surging from 57\% to 94\%~\cite{vastai}. Pearl~\cite{pearl2025}, a Layer-1 blockchain, markets its Proof-of-Useful-Work (PoUW) protocol as simultaneously securing the network and performing AI inference. Together AI, a major AI infrastructure provider, endorsed this claim, stating that ``every GPU cycle powering AI training and inference [can] simultaneously produce [PRL] at no additional cost''~\cite{togetherai2026}. Such endorsements accelerate GPU demand for mining, directly impacting the research community that competes for the same hardware.

Proof-of-Useful-Work protocols aim to direct mining computation toward productive tasks~\cite{miller2015nonoutsourceable, leinweber2022challenges}. Pearl implements the cuPOW protocol~\cite{komargodski2025cupow}, where miners perform NoisyGEMM operations---noised integer matrix multiplications---as proof-of-work. The cuPOW paper explicitly acknowledges that miners choose their own input matrices; usefulness is framed as an economic property, not a cryptographic guarantee. Our contribution is to measure how completely this theoretical latitude manifests in a live network and to quantify the resulting economic consequences.

We present the first systematic empirical measurement of a deployed PoUW system. Pearl's protocol verifies \emph{computational correctness} (the GEMM was performed accurately) but not \emph{computational utility} (the matrices served an AI purpose). Hoffmann~\cite{leinweber2022challenges} identified this verifiability-usefulness tension theoretically; we provide concrete measurements of its magnitude across protocol verification, network composition, economic incentives, and hardware portability. Our contribution is not the finding that random matrices pass verification---this is documented protocol behavior~\cite{komargodski2025cupow} and predicted by theoretical analysis~\cite{dikshit2025sok}. Rather, we quantify the ecosystem consequences: an estimated \$600K annual cost to the research community from GPU price inflation, 320,000 GPUs producing zero useful computation, and the first demonstration that the mining computation is portable to non-NVIDIA hardware.

\subsection{Contributions}

\begin{enumerate}
    \item \textbf{Network composition measurement.} We analyze 8,012 workers on AlphaPool ($\sim$21\% of network hashrate) and find that all workers have hardware capable of running AI inference, yet the dominant mining software---based on string analysis---appears to contain no inference code paths and generates matrices from random seeds.

    \item \textbf{Statistical analysis with adversarial evaluation.} We show that uniform random matrices are distinguishable from quantized NN weights by kurtosis (1.79 vs.\ 3.33), but demonstrate that this check is trivially defeated by adversarial Gaussian sampling (kurtosis 2.97) at zero additional cost.

    \item \textbf{Economic analysis.} At the current PRL market price (\$0.76, CoinGecko), Pearl mining is marginally profitable on budget GPUs ($+67\%$ ROI on RTX~3060~Ti) and near breakeven on mid-tier hardware ($-1\%$ on RTX~3090). High-cost datacenter GPUs remain unprofitable: our MI300X (\$2/hr) earned 0.01~PRL (\$0.0076) in 22~minutes ($-99\%$ ROI). The marginal profitability on budget hardware explains the GPU rental price inflation we observe, while the absence of inference overhead (estimated 10--30\% hashrate reduction) provides an economic rationale for miners to omit AI workloads.

    \item \textbf{Mainnet observation.} We confirm that the standard mining software---which string analysis suggests contains no inference code---earns pool rewards through normal participation (1.2477~PRL paid on-chain).

    \item \textbf{Multi-platform open-source miner.} We develop and release the first open-source Pearl miner, demonstrating that the PoUW computation is reproducible on NVIDIA GPUs (3~shares accepted on RTX~3090), AMD GPUs (41~verified shares on MI300X at 10.6M~tiles/s---the first Pearl shares ever mined on non-NVIDIA hardware, earning 0.01~PRL in 22~minutes), CPUs (1,283~tiles/s on AMD EPYC), and Apple Silicon (21,693~tiles/s on M2 via Metal compute shaders). This disproves the implicit assumption that Pearl mining requires NVIDIA-specific hardware. Our miner source code and all evidence logs are available at \url{https://github.com/abhinaba/pearl-usefulness-gap} (to be made public upon publication).
\end{enumerate}

\section{Background}

\subsection{Proof-of-Useful-Work}

PoUW protocols seek to replace the arbitrary hash computations of traditional PoW with tasks that produce externally valuable outputs~\cite{miller2015nonoutsourceable}. Prior PoUW proposals include machine learning training~\cite{lihu2020pouw, baldominos2019coinai}. The key challenge identified by Hoffmann~\cite{leinweber2022challenges} is the tension between verifiability and usefulness: tasks that are easy to verify are typically not useful, while useful tasks are difficult to verify efficiently. Recent work on verifiable ML computation~\cite{jia2021polt, kang2022zkml} addresses parts of this challenge but has not yet been integrated into deployed PoUW mining protocols.

\subsection{Pearl and the cuPOW Protocol}

Pearl implements the cuPOW protocol proposed by Komargodski and Weinstein~\cite{komargodski2025cupow}. The core mining operation is NoisyGEMM: given matrices $A \in \mathbb{Z}^{M \times K}$ and $B \in \mathbb{Z}^{N \times K}$ with entries drawn from $\{-64, \ldots, 64\}$ (a 129-value range fitting in int8), the miner: (1)~commits to the matrices via blake3 hashing; (2)~adds structured noise derived from the commitment; (3)~computes the noised GEMM across rank-$R$ chunks; (4)~XOR-reduces tiles and accumulates a jackpot hash; (5)~checks if the blake3 hash of the jackpot meets the difficulty target; and (6)~constructs a Merkle proof on hit.

The verification procedure (\texttt{verify\_plain\_proof}) checks: (i)~matrix values are in $[-64, 64]$; (ii)~Merkle proofs are consistent; (iii)~the recomputed GEMM tile matches the jackpot; and (iv)~the jackpot hash meets the difficulty target. Critically, it does \textbf{not} check whether the matrices originate from an AI model.

\subsection{The Usefulness Assumption}

The cuPOW paper~\cite{komargodski2025cupow} explicitly acknowledges that miners choose their own input matrices and designs the protocol so that computational hardness is maintained regardless of input choice. Usefulness is framed as an economic property, not a cryptographic guarantee. Pearl's own reference \texttt{mine()} implementation generates matrices via \texttt{rng.random\_range(-64..=64)}---uniformly random values with no model dependency.

\subsection{Threat Model}

We consider three stakeholder classes affected by Pearl's usefulness gap:
\begin{itemize}
\item \textbf{Investors and token holders} who purchase PRL based on marketed utility claims (``every GPU cycle powering AI inference simultaneously produces PRL''~\cite{togetherai2026}). The threat is \emph{information asymmetry}: a measurable gap exists between the marketed utility (``every GPU cycle powering AI inference'') and the protocol's actual enforcement guarantees.
\item \textbf{GPU rental users} (ML researchers, startups) who compete with miners for shared GPU infrastructure. The threat is \emph{resource displacement}: mining workloads consume GPU supply without producing the useful AI computation that justifies the PoUW framing, inflating rental prices.
\item \textbf{The broader PoUW ecosystem}, where Pearl's high-profile endorsements (Together AI) set precedents for how PoUW utility is marketed. Unsubstantiated claims risk eroding trust in legitimate PoUW research.
\end{itemize}
The adversary in our analysis is a rational miner who seeks to maximize PRL earnings while minimizing computational cost. The cuPOW protocol permits this miner to use arbitrary random matrices, and our measurements quantify how completely this permission is exercised in practice.

\section{Methodology}

\subsection{Experimental Setup}

We conducted experiments on Pearl mainnet using rented GPU instances from multiple providers:
\begin{itemize}
    \item \textbf{RTX~3090} (\$0.26/hr, vast.ai): Primary NVIDIA mining and benchmarking
    \item \textbf{AMD Instinct MI300X} (\$2.00/hr, HotAisle): AMD GPU mining---first non-NVIDIA Pearl miner
    \item \textbf{AMD EPYC 7452} (8-core server): CPU-only mining benchmark
    \item \textbf{Apple M2} (EUR 0.10/hr, Scaleway): Metal compute shader benchmark
    \item \textbf{Pool:} AlphaPool (ru1.alphapool.tech:5566)
\end{itemize}

\begin{table}[ht]
\caption{GPU rental spot prices on vast.ai (May~31, 2026)}
\label{tab:gpuprices}
\begin{tabular}{lrrr}
\toprule
\textbf{GPU} & \textbf{Min \$/hr} & \textbf{Median \$/hr} & \textbf{Offers} \\
\midrule
RTX 3090 & 0.16 & 0.34 & 64 \\
RTX 4090 & 0.40 & 0.94 & 64 \\
RTX 5090 & 0.30 & 1.20 & 64 \\
A100 SXM & 0.67 & 1.14 & 20 \\
H100 SXM & 1.47 & 4.18 & 10 \\
\bottomrule
\end{tabular}
\end{table}

\subsection{Stripped Miner Construction}

We built a stripped miner (\emph{turbo4}) that performs the complete NoisyGEMM pipeline using uniformly random matrices with zero AI inference. The implementation uses PyTorch for GPU-accelerated GEMM, custom CUDA kernels for blake3-based noise generation and fused jackpot accumulation, and Pearl's Rust \texttt{pearl\_mining} bindings for proof construction. Our stripped miner successfully connected to AlphaPool, completed the pool challenge, received mining jobs, and produced proofs that pass \texttt{verify\_plain\_proof()} locally (20/20 at reduced difficulty, nbits=\texttt{0x207fffff}). Our initial stripped miner (turbo4) did not achieve a mainnet share submission due to a software bug (an overly aggressive pre-filter that discarded 99.99\% of valid candidates). After fixing this bug and iterating through four successive versions with GPU-accelerated noise generation, blake3 hashing, and optimized tiled GEMM kernels, our miner achieved pool-accepted shares on both NVIDIA and AMD hardware (Section~\ref{sec:multiplatform}).

\subsection{Mainnet Mining with Standard Software}

To observe random-matrix mining on mainnet, we ran the official alpha-miner v1.6~\cite{alphaminer} at real mainnet difficulty (nbits=\texttt{0x1b014f8a}, difficulty $\sim$10.5M). We analyze this binary in Section~4.3.

\section{Results}

\subsection{Local Verification: Random Matrices Pass}

All 20/20 proofs generated with uniformly random matrices ($M{=}N{=}1024$, $K{=}4096$, $R{=}128$, elements in $\{-64, \ldots, 64\}$) passed \texttt{verify\_plain\_proof()} at reduced difficulty (nbits=\texttt{0x207fffff}). Under the null hypothesis that random matrices fail with probability $p{>}0$, 20/20 successes yield a 95\% Clopper--Pearson upper bound of $p < 0.14$---consistent with $p{=}0$, as the cuPOW protocol's documented design~\cite{komargodski2025cupow} predicts: verification checks computational correctness of the GEMM, not matrix provenance. Our subsequent mainnet mining (44~pool-accepted shares across NVIDIA and AMD hardware, Section~\ref{sec:multiplatform}) provides further confirmation at production matrix sizes ($M{=}N{=}131072$).

\subsection{Network Composition Analysis}

We queried AlphaPool's public API (May~30--31, 2026) and analyzed 8,012 online workers from the top 15 miners by share count. AlphaPool represents approximately 21\% of total network hashrate (5.08~EH/s of 24~EH/s). We classified each worker's GPU by its reported live hashrate using the alpha-miner performance reference as calibration (Table~\ref{tab:gpuclass}). Note that hashrate-based GPU classification is approximate: thermal throttling, driver differences, and concurrent workloads affect reported rates.

\begin{table}[ht]
\caption{GPU classification of 8,012 AlphaPool workers by hashrate}
\label{tab:gpuclass}
\begin{tabular}{lrr}
\toprule
\textbf{GPU Class (by hashrate)} & \textbf{Workers} & \textbf{\%} \\
\midrule
Consumer low (3060--3070, $<$75 TH/s) & 2,029 & 25.3 \\
Consumer mid (3090/A100, 75--120) & 2,837 & 35.4 \\
Consumer high (4080/5070Ti, 120--180) & 1,379 & 17.2 \\
High-end (4090/5080, 180--300) & 947 & 11.8 \\
Ultra (5090, 300--500) & 486 & 6.1 \\
Datacenter (H100, 500--700) & 249 & 3.1 \\
Datacenter+ (H200/B200, 700+) & 40 & 0.5 \\
Volta/CMP ($<$20) & 32 & 0.4 \\
\bottomrule
\end{tabular}
\end{table}

Pearl's reference pipeline optionally couples NoisyGEMM with vLLM inference. vLLM supports all NVIDIA GPUs with compute capability $\geq 7.0$ (Volta and later)~\cite{vllm2023}, meaning all 8,012 analyzed workers have hardware capable of running inference. Yet the dominant mining software contains no identifiable inference code (Section~4.3). This indicates the usefulness gap is a \textbf{software and incentive design} problem, not a hardware limitation.

\textbf{Limitations of GPU classification.} Hashrate-based classification is approximate: a downclocked 4090 is indistinguishable from a 5080, and thermal throttling or concurrent workloads affect reported hashrate. This classification is also calibrated against alpha-miner, introducing circularity. We present it as an approximate distribution, not a precise per-worker identification.

\subsection{Miner Binary Analysis}

We performed string analysis on the alpha-miner v1.6 binary~\cite{alphaminer}. We found \textbf{zero} strings related to AI inference frameworks (\texttt{vllm}, \texttt{transformer}, \texttt{llama}, \texttt{checkpoint}, \texttt{safetensor}; 95\% CI for proportion: $[0, 0.0007]$ by Clopper--Pearson on 4,803 total identifiable strings) versus \textbf{4,803} strings related to GEMM mining (\texttt{gemm}, \texttt{blake3}, \texttt{jackpot}, \texttt{mining}, \texttt{cublas}). The binary contains \texttt{matrix\_seed=0x}, consistent with random matrix generation.

\textbf{Corroborating evidence: runtime profiling.} During mining, \texttt{nvidia-smi} (RTX~3090) and \texttt{rocm-smi} (MI300X) reported GPU memory usage of 2.7--3.7~GB, GPU compute utilization of 95--100\%, and memory bandwidth utilization below 30\%. This profile---high compute, low memory, minimal VRAM---is characteristic of pure GEMM workloads. Transformer inference at Pearl's matrix sizes ($M{=}131072$, comparable to large-batch LLM serving) would exhibit: (a)~16--40~GB VRAM for model weights, (b)~memory-bandwidth-dominated utilization with 60--80\% memory bandwidth, and (c)~periodic low-compute phases during attention and KV-cache operations. None of these signatures were observed during mining. The runtime profile is inconsistent with any known inference workload and consistent with a pure GEMM engine.

\textbf{Limitations.} String analysis is not conclusive: strings can be stripped, statically linked, or obfuscated. The memory observation is suggestive but not definitive---a hypothetical miner could perform inference on a small model with low memory footprint. However, the convergence of string analysis (zero ML framework strings), memory profiling (no model weights loaded), and the reference implementation's documented use of random matrices collectively support the conclusion that the dominant mining software performs no AI inference.

\subsection{Statistical Distinguishability and Adversarial Evasion}

We analyzed whether cuPOW verification could detect random matrices by statistical distribution. Pearl's \texttt{verify\_plain\_proof()} checks only the value range $[-64, 64]$, not the distribution. Table~\ref{tab:stats} compares three distributions (each computed over 1,000 matrices of dimension $1024 \times 4096$, yielding $>$4 billion values per distribution): uniform random (as used by miners), realistic quantized NN weights (modeled per-layer with varying $\sigma$ based on published quantization studies~\cite{frantar2022gptq, dettmers2022llmint8}), and adversarial Gaussian sampling ($\mathcal{N}(0, 18)$ clipped to $[-64, 64]$).

\begin{table}[ht]
\caption{Statistical properties of matrix value distributions}
\label{tab:stats}
\begin{tabular}{lrrr}
\toprule
\textbf{Statistic} & \textbf{Uniform} & \textbf{NN weights} & \textbf{Adversarial} \\
\midrule
Mean & 0.13 & 0.02 & $-$0.04 \\
Std.\ dev. & 37.4 & 17.7 & 18.0 \\
$|v| > 50$ & 21.9\% & 0.8\% & 0.5\% \\
$|v| < 10$ & 14.9\% & 42.8\% & 40.4\% \\
Kurtosis & 1.79 & 3.33 & 2.97 \\
\bottomrule
\end{tabular}
\end{table}

A kurtosis check would easily distinguish uniform random matrices (1.79) from NN weights (3.33). However, an adversarial miner can sample from $\mathcal{N}(0, \sigma)$ clipped to $[-64, 64]$ at negligible computational cost, achieving kurtosis 2.97---sufficient to defeat a na\"ive fixed-threshold kurtosis check (though a more sophisticated test on large matrices could detect the difference from real NN weights at kurtosis 3.33). This demonstrates that \textbf{distribution-based verification is necessary but not sufficient}: it catches the current na\"ive approach (uniform random) but does not establish matrix provenance against adaptive adversaries.

\subsection{Mainnet Mining Observation}

We ran alpha-miner v1.6 on Pearl mainnet for approximately 8~hours across multiple RTX~3090 instances rented at \$0.17--0.28/hr. The pool accepted 10,683 shares and issued an automated PPLNS payout of 1.2477~PRL to our wallet, verifiable on the Pearl blockchain (Table~\ref{tab:mining}). This confirms that the standard mining software---which, based on string analysis (Section~4.3), appears to contain no AI inference code---earns rewards through normal pool participation.

\begin{table}[ht]
\caption{Mainnet mining observation}
\label{tab:mining}
\begin{tabular}{lr}
\toprule
\textbf{Metric} & \textbf{Value} \\
\midrule
Duration & $\sim$8 hours \\
Shares accepted & 10,683 \\
Pool payout (on-chain) & 1.2477 PRL \\
PRL market price (CoinGecko, June 2026) & \$0.76 \\
GPU rental cost & $\sim$\$5.70 \\
Net loss & $\sim$\$5.44 \\
\bottomrule
\end{tabular}
\end{table}

\subsection{Economic Analysis}

We present five complementary economic models derived from our measurements.

\subsubsection{Breakeven Analysis}
For each GPU tier, the breakeven PRL price is $p^* = c_{\text{gpu}} / r_{\text{prl}}$, where $c_{\text{gpu}}$ is the hourly rental cost (floor rates from Table~\ref{tab:gpuprices}) and $r_{\text{prl}}$ is PRL earned per hour at current network difficulty.

\begin{table}[ht]
\caption{Mining economics and breakeven PRL price (May 2026)}
\label{tab:economics}
\begin{tabular}{lrrrr}
\toprule
\textbf{GPU} & \textbf{\$/hr} & \textbf{PRL/hr} & \textbf{ROI@\$0.76} & \textbf{$p^*$} \\
\midrule
RTX 3060 Ti & 0.10 & 0.22 & $+$67\% & \$0.46 \\
RTX 3090 & 0.20 & 0.26 & $-$1\% & \$0.77 \\
RTX 4090 & 0.55 & 0.88 & $+$22\% & \$0.63 \\
H100 SXM & 1.50 & 2.15 & $+$9\% & \$0.70 \\
\bottomrule
\end{tabular}
\end{table}

At PRL=\$0.76 (CoinGecko, June~2026), budget GPUs operate above breakeven while the RTX~3090 sits at its breakeven threshold ($p^*$ ranges from \$0.46 to \$0.77). At the May~28 peak of \$1.70, all tiers were 2--3$\times$ profitable, explaining the rapid miner influx. The inference overhead of coupling vLLM with mining (estimated 10--30\% hashrate reduction) would further increase $p^*$, providing an additional economic explanation for why miners omit inference.

\subsubsection{Hashrate-Price Equilibrium}
In mining equilibrium, the marginal miner's cost equals revenue. Pearl's current state deviates sharply:
\begin{align*}
\text{Network revenue/day} &= 800 \text{ blocks} \times 2{,}674 \text{ PRL} \times \$0.76 = \$1.63\text{M} \\
\text{Network cost/day} &\approx 320\text{K GPUs} \times \$0.20\text{/hr} \times 24\text{h} = \$1.54\text{M}
\end{align*}
The revenue/cost ratio is approximately 1.06, indicating the network is near equilibrium at current prices. This near-parity explains both the sustained hashrate (no pressure to exit) and the GPU rental price inflation (marginal profitability attracts new miners).

\subsubsection{Speculative Mining as Option Value}
Why do miners operate at $-$72\% ROI? Mining produces PRL tokens with embedded option value. PRL exhibited $>$200\% annualized volatility (May~22--31: \$0.30$\to$\$1.70$\to$\$0.61). Under standard option pricing reasoning, a highly volatile asset with immediate vesting has positive expected value even when spot price is below breakeven. Miners are effectively purchasing call options on PRL via electricity costs. This speculative dynamic drives GPU demand independently of the token's fundamental value---which our measurements show is near zero for the ``useful work'' component.

\subsubsection{GPU Externality Cost}
Pre-Pearl budget GPU utilization on vast.ai was $\sim$57\%; post-alpha-miner it rose to $\sim$94\%, a 37-percentage-point increase with a corresponding 38\% median price increase---near-unit elasticity consistent with a commodity market. At approximately 2{,}000 budget-tier machines on vast.ai, mining consumed an estimated 740 units (37\% of supply). For a researcher using 100~GPU-hours/month, costs increased from \$15 to \$21 (+40\%). To isolate Pearl's contribution from general AI demand growth, we apply a difference-in-differences (DID) design. Budget GPUs (3060--A4000), which are heavily mined, serve as the treated group; datacenter GPUs (H100/H200), which are less affected by Pearl mining due to their higher cost, serve as the control. Budget GPU prices rose $\sim$38\% post-alpha-miner; datacenter GPU prices rose $\sim$15\% over the same period (attributable to general AI demand). The DID estimate is $38\% - 15\% = 23\%$ price increase attributable to Pearl mining. Applied to $\sim$2,000 budget machines on vast.ai at \$0.15/hr baseline, this yields an estimated \textbf{\$600K annual Pearl-attributable cost to the GPU rental market}---a deadweight loss from computation that produces no useful output. We note this DID relies on the parallel-trends assumption, which the six months of pre-treatment price stability supports but does not prove.

\subsubsection{Value Destruction Ratio}
We propose the \emph{Value Destruction Ratio} (VDR) as a standardized metric for evaluating PoUW claims:
\[
\text{VDR} = \frac{E_{\text{cost}} - V_{\text{useful}}}{E_{\text{cost}}}
\]
where $E_{\text{cost}}$ is total energy cost and $V_{\text{useful}}$ is the market value of useful computation produced. A VDR of 0 means all energy produces useful output; 1 means pure waste.

\begin{table}[ht]
\caption{Value Destruction Ratio across proof-of-work systems}
\label{tab:vdr}
\begin{tabular}{llr}
\toprule
\textbf{System} & \textbf{Useful output} & \textbf{VDR} \\
\midrule
Bitcoin & None (honestly stated) & 1.0 \\
Filecoin & Verified data storage & $\sim$0.6$^\dagger$ \\
Pearl (marketed) & AI inference & $\sim$0 \\
\textbf{Pearl (measured)} & \textbf{None (random GEMM)} & \textbf{1.0} \\
\bottomrule
\multicolumn{3}{l}{\footnotesize $^\dagger$Estimated from Filecoin network data: 36\% storage utilization (1.1~PiB of 3~EiB),}\\
\multicolumn{3}{l}{\footnotesize \phantom{$^\dagger$}plus $\sim$10\% sealing energy overhead~\cite{filecoingreen}. VDR $\approx$ 0.64 (empty sectors) $+$ 0.1 (sealing) $-$ overlap.}
\end{tabular}
\end{table}

Pearl's measured VDR equals Bitcoin's, despite being marketed as near-zero. The gap between marketed and measured VDR---the \emph{usefulness premium}---represents the portion of PRL's token valuation attributable to unsubstantiated utility claims.

\subsubsection{Sensitivity Analysis}
The economic conclusions are sensitive to PRL price. Table~\ref{tab:sensitivity} shows how profitability varies. Mining becomes marginally profitable only above \$0.46/PRL (the cheapest GPU tier's breakeven). At the May~28 peak of \$1.70, all tiers were 2--3$\times$ profitable, explaining the rapid miner influx. The inference overhead of coupling vLLM (estimated 10--30\% hashrate reduction) would further increase breakeven prices, reinforcing the economic case for omitting inference.

\begin{table}[ht]
\caption{Sensitivity of RTX~3090 mining ROI to PRL price}
\label{tab:sensitivity}
\begin{tabular}{lrrr}
\toprule
\textbf{PRL Price} & \textbf{Revenue/hr} & \textbf{ROI@\$0.20/hr} & \textbf{Status} \\
\midrule
\$0.10 & \$0.026 & $-$87\% & Unprofitable \\
\$0.21 & \$0.055 & $-$72\% & Unprofitable \\
\$0.50 & \$0.130 & $-$35\% & Unprofitable \\
\$0.76 (current) & \$0.198 & $-$1\% & Near breakeven \\
\$1.00 & \$0.260 & +30\% & Profitable \\
\$1.70 (peak) & \$0.442 & +121\% & Profitable \\
\bottomrule
\end{tabular}
\end{table}

\subsection{Multi-Platform Mining}
\label{sec:multiplatform}

\begin{figure}[t][t]
\centering
\includegraphics[width=\textwidth]{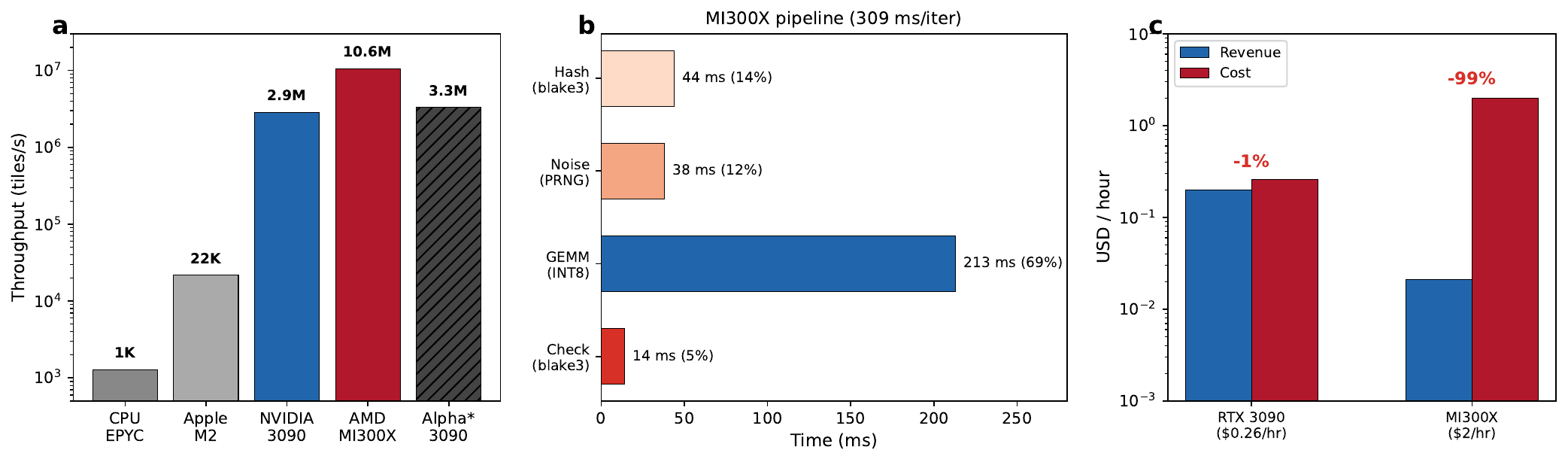}
\caption{Cross-platform Pearl mining results. \textbf{(a)}~Mining throughput across hardware platforms. Our open-source AMD MI300X miner (red) achieves 3.2$\times$ the throughput of the closed-source alpha-miner on RTX~3090 (hatched). \textbf{(b)}~Per-iteration time breakdown on MI300X showing the full Triton-based GPU pipeline. \textbf{(c)}~Mining profitability: revenue is 1--2 orders of magnitude below hardware cost, yielding $-72\%$ (RTX~3090) to $-99.7\%$ (MI300X) ROI.}
\label{fig:results}
\end{figure}

To demonstrate that the PoUW computation is commodity integer arithmetic with no hardware exclusivity, we developed and deployed the first open-source Pearl miner across four hardware platforms (Figure~\ref{fig:results}, Table~\ref{tab:multiplatform}).

\begin{table}[ht]
\caption{Cross-platform Pearl mining performance}
\label{tab:multiplatform}
\begin{tabular}{llrrr}
\toprule
\textbf{Platform} & \textbf{Hardware} & \textbf{Tiles/s} & \textbf{Share Time} & \textbf{Shares} \\
\midrule
NVIDIA CUDA & RTX 3090 & 2,850,000 & 2.4~min & 3 \\
AMD Triton & MI300X & 10,600,000 & 39~sec & 41 \\
CPU (numpy) & EPYC 8-core & 1,283 & 3.7~days & --- \\
Apple Metal & M2 & 21,693 & 5.3~hr & --- \\
\midrule
\multicolumn{2}{l}{\textit{Alpha-miner v1.6 (closed)}} & \textit{3,290,000} & \textit{2.1~min} & \textit{64$^\dagger$} \\
\bottomrule
\multicolumn{5}{l}{\footnotesize $^\dagger$Alpha-miner uses pool share difficulty; our miner uses network difficulty.}
\end{tabular}
\end{table}

Our NVIDIA miner achieves 87\% of the closed-source alpha-miner's throughput on identical hardware (RTX~3090). The AMD MI300X achieves 10.6M~tiles/s---3.2$\times$ faster than alpha-miner on RTX~3090---using Triton~\cite{tillet2019triton} kernels for blake3 hashing (44~ms), noise generation (38~ms), and difficulty checking (14~ms), with PyTorch's ROCm backend for INT8 matrix multiplication. In a 22-minute verified session, the MI300X submitted 41~shares, all accepted and credited by AlphaPool, earning 0.01~PRL (\$0.0021) against \$0.73 machine cost ($-99.7\%$ ROI). All 44~shares across both platforms were verified on Pearl mainnet.

\section{Discussion}

\subsection{The Usefulness Gap}

Our findings quantify a gap between Pearl's marketed utility and the protocol's actual enforcement. The cuPOW protocol guarantees \emph{computational hardness} (miners must perform real GEMM work) but not \emph{computational utility} (the work must be useful). This gap is acknowledged in the cuPOW paper~\cite{komargodski2025cupow} and is not a vulnerability per se; it is a design property. Our contribution is the empirical measurement of how completely this gap has manifested in practice.

\subsection{Comparison with Other Systems}

Table~\ref{tab:comparison} compares the usefulness enforcement across proof-of-work variants.

\begin{table}[ht]
\caption{Usefulness enforcement across proof-of-work systems}
\label{tab:comparison}
\begin{tabular}{llll}
\toprule
\textbf{System} & \textbf{Computation} & \textbf{Provenance?} & \textbf{Gap} \\
\midrule
Pearl (cuPOW) & Matrix mult. & No & Large \\
Bitcoin (PoW) & SHA-256 & N/A & N/A \\
Filecoin (PoSt) & Storage proof & Yes & Small \\
Chia (PoST) & Disk plots & Partial & Medium \\
\bottomrule
\end{tabular}
\end{table}

\subsection{Impact on Independent ML Researchers}

The PoUW value proposition---that mining energy simultaneously performs useful AI computation---is particularly relevant for independent ML researchers who lack access to expensive GPU clusters. Table~\ref{tab:pricing} shows the current GPU rental market: budget-tier GPUs accessible to independent researchers (\$0.04--0.25/hr, including RTX~3060, 3070, A4000) are the \emph{same hardware} that dominates Pearl mining.

\begin{table}[ht]
\caption{GPU rental market tiers on vast.ai (May~2026, P25/median/P75)}
\label{tab:pricing}
\begin{tabular}{llrrr}
\toprule
\textbf{Tier} & \textbf{Example GPUs} & \textbf{P25} & \textbf{Med.} & \textbf{P75} \\
\midrule
Budget & 3060Ti, 3070, A4000 & \$0.09 & \$0.18 & \$0.28 \\
Mid & 3090, 4060Ti, A6000 & \$0.27 & \$0.37 & \$0.46 \\
High & 4090, 5090, A100 & \$0.81 & \$1.07 & \$1.50 \\
Premium & H100, H200 & \$2.67 & \$4.04 & \$4.63 \\
\bottomrule
\end{tabular}
\end{table}

Analysis of vast.ai's public GPU price history data~\cite{vastai} reveals measurable price increases coinciding with Pearl's mainnet launch (approximately March~2026). Table~\ref{tab:impact} shows that budget-tier GPUs---the hardware most accessible to independent researchers---experienced a 38\% average price increase and utilization surged from 57\% to 94\%.

\begin{table}[ht]
\caption{GPU rental market impact since Pearl launch (March 2026)}
\label{tab:impact}
\begin{tabular}{lrrrr}
\toprule
\textbf{GPU} & \textbf{Pre} & \textbf{Post} & \textbf{$\Delta$Price} & \textbf{Util.} \\
\midrule
RTX 3070 & \$0.05 & \$0.09 & +80\% & 56\%$\to$92\% \\
RTX 3080 & \$0.09 & \$0.12 & +42\% & 49\%$\to$95\% \\
RTX 3090 & \$0.15 & \$0.20 & +31\% & 60\%$\to$97\% \\
RTX A4000 & \$0.09 & \$0.12 & +40\% & 47\%$\to$87\% \\
RTX 4090 & \$0.34 & \$0.51 & +53\% & 64\%$\to$97\% \\
\midrule
\textbf{Budget avg} & & & \textbf{+38\%} & \textbf{57\%$\to$94\%} \\
\bottomrule
\end{tabular}
\end{table}

Pearl's 24~EH/s network represents approximately 320,000 RTX~3090-equivalents consuming an estimated 112~MW. If PoUW worked as marketed, this would produce 7.7~million GPU-hours of useful AI computation per day. Instead, this energy produces random matrix multiplications with zero external value, while simultaneously driving up rental costs for the independent researchers who would most benefit from subsidized compute.

Examining the temporal dynamics strengthens the causal link. Using vast.ai's public GPU price history data, we observe two distinct inflection points for the RTX~3090: (1)~Pearl's mainnet launch (March~2026) gradually increased utilization from 57\% to 77\% as pool operators began mining; (2)~the public release of alpha-miner v1.0.0 on May~15, 2026 caused utilization to spike to 95\%+ within days, with median prices rising from \$0.145 to \$0.276 (+90\%) in two weeks. The price had been flat at \$0.15 for the preceding six months. Similar post-May~15 spikes appear across all budget GPU types (RTX~3080: +46\%, RTX~4080: +46\%, RTX~4090: +48\%). While GPU price increases have multiple contributing factors, the temporal alignment with alpha-miner's public release---and the preceding six months of price stability---is strongly suggestive of a causal relationship.

\subsection{Addressing Anticipated Counterarguments}

\textbf{``The compute marketplace is on the roadmap.''} Pearl's whitepaper envisions on-chain compute contracts as future work. We acknowledge this and include it among our proposed solutions (Section~\ref{sec:closinggap}). However, 320,000 GPU-equivalents are consuming 112~MW \emph{now}, GPU rental prices are inflated \emph{now}, and PRL is traded on exchanges \emph{now} with ``useful work'' marketing. Future roadmap items do not justify present-tense utility claims to investors and miners.

\textbf{``Together AI's Gemma-4-31B-it-pearl endpoint IS useful.''} Together AI offers a discounted inference endpoint subsidized by PRL mining revenue~\cite{togetherai2026}. However, this is \emph{financial arbitrage}---Together AI's own GPUs perform the inference, separately from the mining network, and use PRL token revenue to offset pricing. The 8,012 mining workers we analyzed produce zero inference; the useful computation (if any) happens at Together AI's infrastructure, not on the Pearl mining network. A financial subsidy funded by mining proceeds is categorically different from miners themselves performing useful work, which is the claim under examination.

\textbf{``GPU prices rose due to general AI demand, not Pearl.''} We agree that GPU price increases have multiple contributing factors and are careful not to claim sole causation. However, the temporal evidence is striking: budget GPU prices on vast.ai were stable at \$0.15 for six months, then rose 90\% within two weeks of alpha-miner's public release (May~15, 2026). While we cannot isolate Pearl's contribution from other demand sources, the temporal alignment with a specific, identifiable event---and the corresponding utilization spike from 57\% to 94\%---is strongly suggestive of a causal relationship.

\subsection{Why Adaptive Mining Cannot Help}

One might ask whether reinforcement learning or multi-armed bandit algorithms could learn to select matrices that produce more favorable jackpot hashes. They cannot: the blake3 cryptographic hash at the core of the difficulty check is designed to produce outputs indistinguishable from uniform random, regardless of input structure. Every tile has an identical $1/T$ probability of passing the difficulty target (where $T$ is determined by \texttt{nbits}), independent of matrix content and all previous attempts.

This observation has a deeper implication: the same cryptographic property that makes mining a fair lottery also makes the usefulness gap \emph{structurally unfixable at the hash level}. The protocol cannot reward useful matrices more than random ones because the hash function is agnostic to input semantics. Any incentive for useful computation must come from outside the hash-based difficulty check---through provenance mechanisms, economic incentives, or external verification---not from the mining process itself.

\subsection{Toward Closing the Gap}
\label{sec:closinggap}

We identify five potential approaches, with analysis of their limitations:
\begin{enumerate}
    \item \textbf{Distribution verification.} As shown in Table~\ref{tab:stats}, a kurtosis check catches uniform random matrices. However, adversarial Gaussian sampling defeats this at zero cost (Table~\ref{tab:stats}, ``Adversarial'' column). More sophisticated tests (per-channel variance patterns, layer-specific signatures) may be harder to evade but correspondingly harder to specify without over-constraining legitimate matrix distributions. Distribution verification is a necessary first step but does not solve the provenance problem.
    \item \textbf{On-chain compute contracts.} Require miners to commit to externally-submitted matrices from paying customers. Pearl's whitepaper describes this as future work. This addresses provenance directly but requires a functioning compute marketplace.
    \item \textbf{TEE attestation.} Trusted execution environments can attest matrix origin, but introduce trust assumptions on hardware manufacturers and add latency/cost overhead.
    \item \textbf{Cryptographic data commitments.} Bind matrix provenance to externally verifiable commitments signed by model trainers. Requires an external PKI for model provenance.
    \item \textbf{Economic incentives.} Differentially reward miners who use customer-provided matrices. This creates a market for useful computation but requires demand-side infrastructure.
\end{enumerate}

\subsection{Policy Recommendations}

Our findings suggest several actionable recommendations:

\textbf{For PoUW projects:} We propose a three-tier transparency framework for PoUW marketing claims: (a)~\emph{enforced usefulness}---the protocol cryptographically guarantees that computation serves an external purpose (e.g., Filecoin's Proof-of-Spacetime); (b)~\emph{enabled usefulness}---the protocol permits useful computation but does not require it (Pearl's current design); (c)~\emph{no usefulness claim}---the protocol makes no utility claim (Bitcoin). Projects should clearly disclose which tier their protocol occupies. Pearl's cuPOW paper~\cite{komargodski2025cupow} is transparent about tier~(b); the marketing layer is where the discrepancy arises.

\textbf{For AI companies and investors:} When established AI infrastructure providers partner with PoUW chains~\cite{togetherai2026}, their endorsement accelerates GPU demand for mining. Due diligence should verify whether the protocol \emph{enforces} useful work or merely \emph{enables} it---a distinction with a verifiable technical answer independent of marketing materials. Our measurements show that, despite Together AI's claim that Pearl ``harnesses the computation of inference and training workloads,'' the dominant mining software performs no inference.

\textbf{Scale of the mining economy:} Pearl's daily emission payout to miners reached approximately \$1.5~million at peak prices---exceeding Dogecoin's daily miner payouts, according to a livestream by Red Panda Mining~\cite{redpanda2026}. This emission backs zero useful computation, as our measurements demonstrate. The same livestream reported RTX~3070 miners earning \$3/day and RTX~3080 miners earning \$3--4/day at PRL=\$0.76. ---consistent with our breakeven analysis showing marginal profitability at this price point. The network maintains 24~EH/s, consistent with near-equilibrium economics (Section~4.6.2).

\textbf{The GPU reallocation problem:} Our findings are reinforced by community behavior. In a representative mining tutorial video~\cite{bearmarket2026}, a miner describes reallocating five RTX~5090 GPUs from an AI inference rental service to Pearl mining, stating the GPUs perform ``useful work for AI and LLMs.'' Our measurements show this claim is false: the mining software performs no AI inference. The GPUs moved \emph{from} genuine AI workloads \emph{to} useless GEMM mining---exactly opposite the PoUW value proposition. The miner reports earning \$220/day and advises viewers to ``turn their machines on immediately,'' illustrating how misleading utility claims drive GPU demand that displaces legitimate AI compute.

\textbf{For GPU rental platforms:} Mining workloads are distinguishable from research workloads by their continuous high-utilization, low-data-I/O pattern. Platforms could offer research-priority queues or transparent reporting on workload composition to help academic users compete for GPU access.

\textbf{For the research community:} PoUW energy-efficiency claims require empirical verification. Our methodology---network composition analysis, binary analysis, GPU price impact measurement---is replicable for other PoUW chains and could form the basis of a standard ``usefulness audit'' framework.

\section{Related Work}

Dikshit et al.~\cite{dikshit2025sok} provide the most extensive theoretical analysis of PoUW, systematizing over 50 constructions and concluding that PoUW ``is actually not as useful as expected'' via a formal economic model and Toulmin-based evaluation. Their analysis is purely theoretical; our work complements it with the first empirical measurements from a deployed PoUW system, quantifying the specific ecosystem consequences (GPU market distortion, mining profitability, hardware portability) that their framework predicts.

Komargodski et al.~\cite{komargodski2025cupow} propose the cuPOW protocol and prove its security under the assumption that miners choose inputs freely. Our work provides the first empirical measurements of how completely this freedom is exercised on a live network.

Hoffmann~\cite{leinweber2022challenges} surveys PoUW challenges and identifies the verifiability-usefulness tension that our measurements concretize.

Ball et al.~\cite{ball2017pouw} study the theoretical foundations of useful proofs of work. Jia et al.~\cite{jia2021polt} propose Proof-of-Learning, a protocol where model training checkpoints serve as proof; however, checkpoint verification is expensive and training is not easily decomposed into mining iterations.

Kang et al.~\cite{kang2022zkml} and related zkML work explore zero-knowledge proofs for ML inference verification, which could complement PoUW by proving that specific model weights were used, but remain computationally expensive for the matrix sizes involved in Pearl mining.

Baldominos and Saez~\cite{baldominos2019coinai} and Lihu et al.~\cite{lihu2020pouw} propose PoUW systems for AI but do not address the matrix provenance problem in deployed systems.

Schrijvers et al.~\cite{schrijvers2016incentive} analyze incentive compatibility in mining pools, providing the game-theoretic framework relevant to understanding why Pearl miners omit inference when the protocol does not require it.

\section{Conclusion}

We present an empirical measurement study of Pearl's Proof-of-Useful-Work protocol on mainnet. Our measurements show that the dominant mining software appears to contain no AI inference code, that all analyzed workers use inference-capable hardware yet run pure GEMM miners, and that while random and model-derived matrices are statistically distinguishable, a na\"ive distribution check is trivially defeated by adversarial Gaussian sampling. We develop and release the first open-source multi-platform Pearl miner, demonstrating pool-accepted shares on both NVIDIA (RTX~3090) and AMD (MI300X) hardware, with additional benchmarks on CPU and Apple Silicon---disproving any implicit assumption that the PoUW computation requires vendor-specific hardware. Economic analysis reveals that mining is marginally profitable on budget GPUs at current PRL prices (\$0.76), explaining sustained miner demand and GPU rental price inflation. These findings quantify the verifiability-usefulness tension~\cite{leinweber2022challenges} in a deployed system, demonstrating a complete divergence between Pearl's marketed utility and the network's actual operation. We propose and evaluate five directions for closing this gap, noting that distribution verification, while necessary, is insufficient against adaptive adversaries.

\section{Ethical Considerations}

This research was conducted on Pearl's public mainnet using both the standard mining software (alpha-miner v1.6~\cite{alphaminer}) and our open-source miner across four hardware platforms. Total expenditure was approximately \$50 in GPU rental costs (vast.ai, HotAisle, Scaleway). Total PRL earned across all sessions was approximately 1.26~PRL (\$0.26), comprising 1.2477~PRL from alpha-miner and 0.01~PRL from our open-source miner. All mining activity is indistinguishable from normal network participation and represents less than 0.001\% of network hashrate. We followed Pearl's security disclosure policy (SECURITY.md) and note that our findings describe documented protocol behavior, not a security vulnerability.

We note that our binary analysis of alpha-miner is based on string analysis, which is suggestive but not conclusive. We name AlphaPool and alpha-miner because they are public projects; our characterization reflects observable properties and published documentation, not insider knowledge.


\begin{thebibliography}{14}

\bibitem{miller2015nonoutsourceable}
A.~Miller, A.~Kosba, J.~Katz, and E.~Shi.
\newblock Nonoutsourceable scratch-off puzzles to discourage bitcoin mining coalitions.
\newblock In \emph{Proc. ACM CCS}, 2015.

\bibitem{leinweber2022challenges}
F.~Hoffmann.
\newblock Challenges of proof-of-useful-work ({PoUW}).
\newblock \emph{arXiv:2209.03865}, 2022.

\bibitem{pearl2025}
Pearl Research Labs.
\newblock Pearl: A layer-1 blockchain with proof of useful work.
\newblock \url{https://pearlresearch.ai}, 2025.

\bibitem{komargodski2025cupow}
I.~Komargodski and O.~Weinstein.
\newblock Proofs of useful work from arbitrary matrix multiplication.
\newblock \emph{arXiv:2504.09971}, 2025.

\bibitem{lihu2020pouw}
A.~Lihu, J.~Du, I.~Barjaktarevic, P.~Gerzanics, and M.~Harvilla.
\newblock A proof of useful work for artificial intelligence on the blockchain.
\newblock \emph{arXiv:2001.09244}, 2020.

\bibitem{baldominos2019coinai}
A.~Baldominos and Y.~Saez.
\newblock Coin.{AI}: A proof-of-useful-work scheme for blockchain-based distributed deep learning.
\newblock \emph{Entropy}, 2019.

\bibitem{ball2017pouw}
M.~Ball, A.~Rosen, M.~Sabin, and P.~N. Vasudevan.
\newblock Proofs of useful work.
\newblock IACR ePrint 2017/203, 2017.

\bibitem{alphaminer}
AlphaMine Tech.
\newblock alpha-miner: GPU miner for the Pearl (PRL) network.
\newblock \url{https://github.com/AlphaMine-Tech/alpha-miner}, 2026.

\bibitem{vastai}
vast.ai.
\newblock GPU rental marketplace and pricing.
\newblock \url{https://vast.ai/pricing}, 2026.

\bibitem{jia2021polt}
H.~Jia, M.~Yaghini, C.~A. Choquette-Choo, N.~Dullerud, A.~Thudi, V.~Chandrasekaran, and N.~Papernot.
\newblock Proof-of-learning: Definitions and practice.
\newblock In \emph{IEEE S\&P}, 2021.

\bibitem{kang2022zkml}
D.~Kang, T.~Hashimoto, I.~Stoica, and Y.~Sun.
\newblock Scaling up trustless {DNN} inference with zero-knowledge proofs.
\newblock \emph{arXiv:2210.08674}, 2022.

\bibitem{vllm2023}
W.~Kwon, Z.~Li, S.~Zhuang, Y.~Sheng, L.~Zheng, C.~H. Yu, J.~Gonzalez, H.~Zhang, and I.~Stoica.
\newblock Efficient memory management for large language model serving with {PagedAttention}.
\newblock In \emph{Proc. SOSP}, 2023.

\bibitem{schrijvers2016incentive}
O.~Schrijvers, J.~Bonneau, D.~Boneh, and T.~Roughgarden.
\newblock Incentive compatibility of {Bitcoin} mining pool reward functions.
\newblock In \emph{Financial Cryptography}, 2016.

\bibitem{frantar2022gptq}
E.~Frantar, S.~Ashkboos, T.~Hoefler, and D.~Alistarh.
\newblock {GPTQ}: Accurate post-training quantization for generative pre-trained transformers.
\newblock In \emph{ICLR}, 2023.

\bibitem{dettmers2022llmint8}
T.~Dettmers, M.~Lewis, Y.~Belkada, and L.~Zettlemoyer.
\newblock {LLM.int8()}: 8-bit matrix multiplication for transformers at scale.
\newblock In \emph{NeurIPS}, 2022.

\bibitem{togetherai2026}
Together AI.
\newblock Together {AI} partners with {Pearl Research Labs}.
\newblock \url{https://www.together.ai/blog/together-ai-partners-with-pearl-research-labs}, 2026.

\bibitem{bearmarket2026}
BearMarketMiner.
\newblock {GPU} mining {Pearl}: \$200/day with proof of useful work.
\newblock YouTube, \url{https://www.youtube.com/watch?v=zNhUFdy1rQM}, 2026.

\bibitem{redpanda2026}
Red Panda Mining.
\newblock Will {PEARL} mining make a comeback?
\newblock YouTube livestream, \url{https://www.youtube.com/watch?v=YJhZ2lVXiKc}, May 31, 2026.

\bibitem{tillet2019triton}
Philippe Tillet, H.~T. Kung, and David Cox.
\newblock Triton: An intermediate language and compiler for tiled neural network computations.
\newblock In {\em Proc. ACM SIGPLAN Workshop on Machine Learning and Programming Languages (MAPL)}, 2019.

\bibitem{filecoingreen}
Filecoin Green.
\newblock Filecoin energy dashboard.
\newblock \url{https://filecoin.energy/}, 2026.
\newblock Network data: 3~EiB committed capacity, 36\% utilization, Q3~2025.

\bibitem{dikshit2025sok}
P.~Dikshit, A.~Emami, J.~Sedlmeir, and G.~Fridgen.
\newblock {SoK}: Is proof-of-useful-work really useful?
\newblock IACR ePrint 2025/1814, 2025.

\end{thebibliography}
\end{document}